\documentclass[a4paper,11pt]{amsart}
\usepackage{graphicx}
\usepackage{amssymb}
\begin{document}

\hyphenation{gra-vi-ta-tio-nal re-la-ti-vi-ty Gaus-sian
re-fe-ren-ce re-la-ti-ve gra-vi-ta-tion Schwarz-schild
ac-cor-dingly gra-vi-ta-tio-nal-ly re-la-ti-vi-stic pro-du-cing
de-ri-va-ti-ve ge-ne-ral ex-pli-citly des-cri-bed ma-the-ma-ti-cal
de-si-gnan-do-si coe-ren-za pro-blem gra-vi-ta-ting geo-de-sic
per-ga-mon cos-mo-lo-gi-cal gra-vity cor-res-pon-ding
de-fi-ni-tion phy-si-ka-li-schen ma-the-ma-ti-sches ge-ra-de
Sze-keres con-si-de-red tra-vel-ling ma-ni-fold re-fe-ren-ces
geo-me-tri-cal in-su-pe-rable sup-po-sedly at-tri-bu-table
Bild-raum in-fi-ni-tely counter-ba-lan-ces iso-tro-pi-cally
pseudo-Rieman-nian cha-rac-te-ristic geo-de-sics
Koordinaten-sy-stems ne-ces-sary Col-la-bo-ra-tion ine-qua-li-ties
coa-le-scence}

\title[On the LIGO-VIRGO search of coalescing-binary signals] {{\bf On the LIGO-VIRGO search\\of coalescing-binary signals}}

\author[Angelo Loinger]{Angelo Loinger}
\address{A.L. -- Dipartimento di Fisica, Universit\`a di Milano, Via
Celoria, 16 - 20133 Milano (Italy)}
\author[Tiziana Marsico]{Tiziana Marsico}
\address{T.M. -- Liceo Classico ``G. Berchet'', Via della Commenda, 26 - 20122 Milano (Italy)}
\email{angelo.loinger@mi.infn.it} \email{martiz64@libero.it}

\vskip0.50cm

\begin{abstract}
No signal of the arrive of gravitational waves has been recorded
by LIGO-VIRGO laser-interferometric detectors during recent
science runs. We give a theoretical explanation of this fact.
\end{abstract}

\maketitle

\normalsize

\vskip1.20cm \noindent \textbf{Summary} -- \emph{Introduction}:
themes and main results. -- \textbf{1, 1bis.} Kundt's ideas on the
origin of the gamma-ray bursts. -- \textbf{2.} The low-latency
search of coalescing-binary signals performed by the LIGO-VIRGO
gravitational-wave detectors. The post-Newtonian approximation of
general relativity, the numerical-relativity simulations, and the
role of the gravitational energy-pseudotensor. --
\textbf{3a}$\div$\textbf{3k}. Eleven arguments concerning the
gravitational waves.

\vskip1.20cm \noindent \emph{\textbf{Introduction}} --It is a
widespread conviction that by virtue of the emission of
gravitational waves (GWs) the orbital evolution of binary neutron
stars (NS-NS) must end in a merger of the two components, see
\emph{e.g.} papers \cite{1}, \cite{2}. The coalescence mechanisms
would give origin to bursts of GWs and to gamma-ray bursts (GRBs).
In the abstract of the paper by LIGO-VIRGO Coll. \cite{1} we find
the following sentences: ``Over the course of the science run,
three gravitational-wave triggers passed all of the low-latency
selection cuts. Of these, one was followed up by several of our
observational partners. Analysis of the gravitational-wave data
leads to an estimated false alarm rate of once every 6.4 days,
falling far short of the requirement for a detection based solely
on gravitational-wave data.''

\par Papers \cite{2} are theoretical. Blanchet's paper develops
the general theory of the post-Newtonian (PN) approximation, and
the application to the problem of two mass-points.

\par We shall prove that from a rigorous general-relativistic
standpoint the belief of an inspiral orbital evolution with a
final merger, generated by emission of GWs, is unfounded. This
belief had \emph{in primis} its origin in the \emph{observed}
shrinking of the orbits of PSRB1913+16, a shrinking whose actual
causes are different from the emission of GWs, see paper \cite{3}.
The pulsar of this binary is a ``recycled'' star, \emph{i.e.} an
object that was spun up by material accretion. Of course, in
general an accretion from the circumstellar medium (CSM) can
originate an energy-dissipation mechanism, which can give an
inspiral evolution of the orbits and a final coalescence of the
two members of a binary.

\par An obvious corollary of our critical remarks (see sects.
\textbf{2}, \textbf{3}) is that the alternative Kundt's
interpretation of the origin of the GRBs must be seriously
considered \cite{4}. We give now a summary of the pertinent ideas
of this author.

\vskip1.20cm \noindent \textbf{1.} -- According to Kundt \cite{4},
\emph{all} kinds of (non-terrestrial) GRBs are emitted by the
surfaces of isolated neutron stars \emph{of our galaxy}, at
distances $d$ within $10^{-2} \leq d/ \textrm{kpc} \leq 0.5$.
These stars belong to the class of the \emph{magnetars} (whose
surface magnetic-field strengths $B$ are of the order $10^{15}
\textrm{G}$), which are re-interpreted by Kundt as
``\emph{throttled}'' \emph{pulsars}. He emphasizes (see in
\cite{4} his seminar in Washington, 2007) that the magnetosphere
of these pulsars ``is deeply indented by a low-mass accretion disk
assembled from its CSM $[$circumstellar medium$]$. These disks
tend to be perpendicular to the Milky Way. Their (anisotropic)
emission -- by ricocheting, accreting ``blades'' -- peak near
their disk plane, strengthening an isotropic appearance of the
bursts in the sky. -- The afterglows are light echoes, or
transient reflection nebulae. -- The short GRBs, of peak duration
$<2$ sec, result by accretion of a single blob (blade), of size of
a terrestrial mountain: they are modulated by the throttled
pulsar's spin (of period $5s$ to $10s$), and soften and tail off
within some $10^{2}$ sec. -- The long GRBs are superposition of
short GRBs, cf. the July 1994 accretion by Jupiter of comet
Shoemaker-Levy. -- Occasionally, accretion onto a throttled pulsar
can trigger additional high-energy activity, of much longer
duration (than $10^{2}$ sec).''

\par Kundt explains that \emph{relativistic} red-shifts can be
generated also by nearby \emph{Galactic} neutron stars.

\vskip1.20cm \noindent \textbf{1bis.} -- Kundt's specification of
 \emph{non-terrestrial} GRBs has also the aim to emphasize that
they ``do not necessarily require exotic sources'' \cite{5}. The
ionosphere is the positive pole of the terrestrial capacitor $C$,
whose negative pole is Earth's surface. The permanent charge
amounts to $(0.6\pm 0.1)$MV. The terrestrial GRBs are simply
discharges of $C$.

\vskip1.20cm \noindent \textbf{2.} -- A low-latency search of
coalescing-binary signals has been performed by LIGO and VIRGO
GW-detectors during their $S6$ and $VSR2$ science runs \cite{1}.
The mass range of the binary components was restricted between $1$
and $34$ solar masses, with a total mass of the system between $2$
and $35$ solar masses. A significant aim was that to allow a
detection of electromagnetic counterparts (short, hard gamma-ray
bursts) to GW-candidates.

\par A set of mathematical GW-forms (``templates'') had been
computed starting from the star motions in the second
post-Newtonian approximation. The search aimed to compare these
templates with the output interferometric signals. The efficiency
of the apparatuses was verified by ``hard'' and ``soft''
injections of artificial signals. --

\par Blanchet \cite{2} starts from the Einsteinian field
equations written in harmonic coordinates $y^{\mu},
(\mu=0,1,2,3)$, characterized by the four conditions

\begin{equation} \label{eq:one}
\frac{\partial}{\partial y^{\mu}}\left( \sqrt{-g} \,
g^{\mu\nu}\right) = 0 \quad;
\end{equation}

then, he considers a Minkowskian \emph{Bildraum} \textbf{B} in
which the following approximate relations hold:

\begin{equation} \label{eq:two}
\sqrt{-g} \, g^{\mu\nu} \approx  h^{\mu\nu} + \eta^{\mu\nu} \quad,
\end{equation}

where $\eta^{\mu\nu}$ is the customary Minkowski tensor (spatial
Cartesian orthogonal coordinate $x^{1}, x^{2},x^{3})$, and
$h^{\mu\nu} (x^{0},x^{1}, x^{2},x^{3})$ is a symmetric field
acting in the flat spacetime \textbf{B}. This field satisfies the
equations

\begin{equation} \label{eq:three} \frac {\partial h^{\mu \nu}}
{\partial x^{\mu} }= 0  \quad,
\end{equation}

\begin{equation} \label{eq:four}
\eta^{\alpha \beta} \, \frac{\partial^{2} h^{\mu \nu}}{\partial
x^{\alpha} \, \partial x^{\beta}} = \frac{16\pi G}{c^{4}}\,
\tau^{\mu\nu} \quad;
\end{equation}

$G$ is the gravitational constant, and $\tau^{\mu\nu}$ is defined
by the following equalities:

\begin{equation} \label{eq:five}
\tau^{\mu\nu} \equiv |g|  \, T^{\mu\nu} +  \frac{c^{4}}{16\pi G}
\, A^{\mu  \nu}\quad ,
\end{equation}

where $T^{\mu\nu}$ is the matter tensor and $A^{\mu \nu}$ is a
\emph{pseudo}-tensor, which behaves as a true tensor with respect
to Lorentz transformations in \textbf{B} of the coordinates
$x^{0},x^{1}, x^{2},x^{3}$; the explicit form of $A^{\mu\nu}$ is
given by Blanchet's eqs. $(11)$. According to a widespread
conviction, $A^{\mu\nu}$ represents the energy-momentum-stress
tensor of the gravitational field $h^{\mu\nu}$.

\par The post-Newtonian expansion is of the form (``near zone''):

\begin{equation} \label{eq:six}
h^{\mu \nu} (\textbf{\emph{x}}
, t, c) = \sum_{n=2}^{+\infty} \,
\frac{1}{c^{n}} \, \underset{n}{h}^{\mu  \nu} (\textbf{\emph{x}},
t, \ln c)\quad ,
\end{equation}

\begin{equation} \label{eq:seven}
\tau^{\mu \nu} (\textbf{\emph{x}}, t, c) = \sum_{n=-2}^{+\infty}
 \, \frac{1}{c^{n}} \, \underset{n}{\tau}^{\mu
\nu} (\textbf{\emph{x}}, t, \ln c) \quad;
\end{equation}

the order $nPN$ corresponds to the order $(v/c)^{2n}$ in the
equations of motion, where $v$ is the orbital velocity.

\par The mentioned pseudo-tensor $A^{\mu \nu}$ is essentially the
approximate form of the pseudo-tensor $L^{\mu \nu}$ of the
\emph{exact} Einstein theory in harmonic coordinates (see
\emph{e.g.}, Fock \cite{6}). $L^{\mu \nu}$ is different from the
standard pseudo-tensor and from Landau and Lifshitz's
pseudo-tensor \cite{7}. Indeed, this notion does not possess a
unique mathematical expression -- \emph{et pour cause}, as we
shall now demonstrate.

\par With reference to the standard pseudo-tensor $t^{\mu
\nu}$, Bauer showed that in a Minkowskian spacetime there exist
systems of coordinates for which $t^{\mu \nu}$ is different from
zero \cite{8}; a statement which is true for any form of
gravitational pseudo-tensor. Bauer gave this example; if $r,
\vartheta, \varphi$ are the usual polar coordinates in a
Minkowskian manifold, the metric of the following spatial
coordinates

\begin{equation}\label{eq:eigth}
\xi =  \frac{1}{3} \, r^{3} \quad ; \quad \eta = -\cos \vartheta
\quad ; \quad \zeta = \varphi
\end{equation}

yields a $t^{\mu \nu}$ different from zero.  It is a curious fact
that these coordinates coincide with the initially chosen
coordinates by Schwarzschild in his first fundamental memoir of
1916 \cite{9}. (Schwarzschild's original solution holds,
mathematically and physically, in the \emph{entire} spacetime,
with the only exception of the origin $r=0$, seat of the
point-mass which generates the investigated gravitational field).

\par Bauers's criticism is sufficient to invalidate the physical
meaning of the gravitational pseudo-tensors: if a simple
coordinate change in a Minkowskian manifold like that of eqs.
(\ref{eq:eigth}) can alter the values of the pseudo-tensors, we
can affirm that the shrinking of the stellar orbits due to the
emission of GWs, according to the post-Newtonian approximation
(\emph{e.g.}), is quite illusive. And also quite illusive is the
existence of a gravitational radiation-reaction force, which
appears at the $2.5$ PN order.

\par But a decisive criticism is as follows: the post-Newtonian
method treats the neutron stars of a binary as material points of
a ``dust''; consequently, the orbits are \emph{geodesic} lines
with \emph{no} emission of GWs. In sect. \textbf{3} we shall
return to these motions in a more detailed way. --

\par Since PN-method is an approximate procedure valid for
\emph{slow} motions and \emph{weak} gravitational fields, which is
based on expansions with respect to a parameter $\varepsilon \sim
(v/c)^{2} \sim GM/(c^{2}r)$ -- where $M$ is the total mass of the
considered NS-NS binary, and $r$ the separation between its
components --, people thought that PN-method is not adequate in
proximity of the coalescence. Thus, many
\emph{numerical-relativity} (NR) \emph{simulations} have been
performed by various authors to describe this hypothesized last
stage; see, \emph{e.g.}, the first two papers quoted in \cite{2}.
They concern the orbital evolutions of a binary composed of a
white-dwarf star and a neutron star, and of a binary composed of
two neutron stars, respectively. Both papers treat \emph{extended}
objects, whose hydrodynamical structure is characterized by
suitable equations of state (EOSs) of various polytropic kinds. Of
course, without an \emph{existence theorem} any numerical
computation is based on the belief that it can yield a reasonably
approximate representation of the exact solution. In the present
instances, the computations lack also of observational
confirmations of their results. And the fact that the
gravitational pseudo-tensor is a true tensor only with respect to
linear coordinate transformations tells us that -- as for the
post-Newtonian method -- the obtained solutions have a physically
deceptive character. --

\par In the following sect. \textbf{3} we give a r\'esum\'e of the
main arguments against the reality of \emph{physical} GWs.

\vskip1.20cm \noindent\textbf{3a.} -- The results of the
computations which involve the gravitational
energy-momentum-stress pseudotensor \emph{depend on the chosen
reference frame}. This means that they cannot have a
general-relativistic invariant character, and therefore a true
physical meaning. Example: the gravitational shrinkings of stellar
orbits. --

\vskip0.60cm \noindent\textbf{3b.} -- \emph{The proper concept of
point-mass in} GR can be obtained as follows. In his second
fundamental memoir of 1916 \cite{10} Schwarzschild solved the
problem of the gravitational field created by a sphere of an
incompressible and homogeneous fluid. In the last section he gave
some hints for deriving from the field of the sphere the
gravitational field of a mass-point, which had been directly
deduced by him in his previous basic work \cite{9}. The suggested
procedure is not trivial, because externally to the sphere
Schwarzschild's radial coordinate is equal to
$(r^{3}+\varrho)^{1/3}$, and the constant $\varrho$ is different
from the value $(2GM/c^{2})^{3}$, which it has in the
corresponding points of the field generated by a point-mass $M$.
The cause of this difference is the following: Schwarzschild
postulated -- in accordance with the classical gravitational
theory -- that at the surface of the sphere (of mass $M$) there is
a coincidence between the internal and external values of the
metric tensor \emph{and of its first derivatives}. If he had
assumed a coincidence only for the values of the metric tensor (as
Weyl did, in conformity with the procedure of elasticity theory),
he would have found $\varrho = (2GM/c^{2})^{3}$. In the paper
\cite{11} the reader finds an explicit development of
Schwarzschild's suggestion. --

\vskip0.60cm \noindent\textbf{3c.} -- In 1926 Levi-Civita
\cite{12} gave a \emph{geometrically explicit} explanation of the
general form of solution (de Sitter, Eddington) to the
Schwarzschild problem to find the Einsteinian field created by a
point-mass $M$ at rest. He adopted a Palatini's procedure
\cite{13}, which yields \emph{the appropriate geometrical
definition of spherical symmetry in a curved spatial manifold},
and the justification of the employment in it of the polar
coordinates $r\, (\geq 0), \vartheta \, (0\leq \vartheta \leq
\pi), \varphi \, (0\leq \ \varphi < 2\pi)$. He founds, with de
Sitter and Eddington:

\begin{equation} \label{eq:nine}
\textrm{d}s^{2} = \left[ 1 - \frac{2m}{R(r)}\right] \,
c^{2}\textrm{d}t^{2} -  \left[ 1 - \frac{2m}{R(r)}\right]^{-1}
\left[ \textrm{d}R(r) \right]^{2} - \left[ R(r) \right]^{2}
(\textrm{d}\vartheta^{2} + \sin^{2}\vartheta
\textrm{d}\varphi^{2}) \quad ,
\end{equation}

where $m\equiv GM/c^{2}$, and $R(r)$ is any regular function of
$r$, which gives a Minkowskian $\textrm{d}s^{2}$ at $r=\infty$.
For $R(r)=r$ we have the standard (Hilbert-Droste-Weyl) form of
solution; for $R(r) = \left[ r^{2}+(2m)^{3} \right]^{1/3}$ and
$R(r) = r+2m$ the original Schwarzschild \cite{9} and Brillouin's
\cite{14} forms of solution, respectively. It is evident from
Levi-Civita's treatment that eq. (\ref{eq:nine}) has a
mathematical and physical meaning \emph{only for} $R(r) > 2m$, and
that \emph{no} role inversion between $R(r)$ and $t$ for $R(r)
\leq 2m$ is allowed.

\par \emph{Tenporarily forgetting} that when $R(r)
\leq 2m$, metric (\ref{eq:nine}) loses any meaning, we can claim
that the surface area $A=4\pi (2m)^{2}$ represents an invariant
and significant notion -- and therefore that the so-called
``Schwarzschild radius'' $2m$ is physically meaningful. But this
forgetting is not permitted, and thus we understand why the
Founding Fathers of GR rejected the notion of BH. As a matter of
fact, the astrophysical phenomena that have been interpreted as
originated by a BH can be plainly interpreted as due to a great,
or enormous, mass concentrated in a relatively small space region.
In particular, no hypothesized property of the ``event horizons''
has ever been observed. (Remark that radial test-particles and
radial light-rays arrive at $R(r)=2m$ with \emph{zero} velocity
and \emph{zero} acceleration). Kundt thinks that the stellar-mass
BH-candidates are in reality neutron stars inside massive
accretion disks, and that the central engine of an AGN is a
nuclear-burning disk. --

\vskip0.60cm \noindent\textbf{3d.} -- It follows from sects.
 \textbf{3a}, \textbf{3b}, \textbf{3c}, that GR does not need the
 notion of mass renormalization. (In the \emph{classical}
 electromagnetic theory there is no charge renormalization). Under
 this respect, Einstein theory is quite analogous to Newton
 theory: ``bare'' mass means zero mass. And the notion of
 ``ADM-mass''must be discarded. --

\vskip0.60cm \noindent\textbf{3e.} -- Let us consider a system of
non-interacting bodies moving in a Minkowskian spacetime. If
$q^{\mu}(\tau)$, $\mu=0,1,2,3)$, are the translational coordinates
of one of them as functions of proper time $\tau$, we have that

\begin{equation} \label{eq:ten}
\mathcal{L}_{(0)} := \eta_{\mu\nu} \, \frac{\textrm{d}q^{\mu}}
{\textrm{d}\tau} \, \frac{\textrm{d}q^{\nu}} {\textrm{d}\tau} =
c^{2}
\end{equation}

is a first integral of Lagrange equations

\begin{equation} \label{eq:eleven}
\frac{\partial \mathcal{L}_{(0)}}{\partial q^{\mu}} - \,
\frac{\partial}{\partial \tau} \, \,  \frac{\partial L_{(0)}}
{\partial \, (\textrm{d}q^{\mu} / \textrm{d}\tau)} = 0 \quad,
\end{equation}

from which:

\begin{equation}\label{eq:twelve}
\frac{\textrm{d}^{2}q^{\mu}} {\textrm{d}\tau^{2}} = 0 \quad ,
\end{equation}

\emph{i.e.} a rectilinear and uniform motion.

\par Quite analogously, if we consider a system of bodies
interacting \emph{only} gravitationally and moving in the
Riemann-Einstein manifold created by them, we have that

\begin{equation} \label{eq:thirteen}
\mathcal{L} := g_{\mu\nu}\left[ q(\tau) \right] \,
\frac{\textrm{d}q^{\mu}} {\textrm{d}\tau} \,
\frac{\textrm{d}q^{\nu}} {\textrm{d}\tau} = c^{2}
\end{equation}

is a first integral of Lagrange equations

\begin{equation} \label{eq:fourteen}
\frac{\partial \mathcal{L}}{\partial q^{\mu}} - \,
\frac{\partial}{\partial \tau} \, \,  \frac{\partial L} {\partial
\,  (\textrm{d}q^{\mu} / \textrm{d}\tau)} = 0 \quad,
\end{equation}

which coincides with the \emph{geodesic} equations

\begin{equation} \label{eq:fifteen}
\frac{\textrm{d}^{2}q^{\mu}} {\textrm{d}\tau^{2}} +
\Gamma^{\mu}_{\,\, \varrho \sigma}\, \frac{\textrm{d}q^{\varrho}}
{\textrm{d}\tau} \, \frac{\textrm{d}q^{\sigma}} {\textrm{d}\tau} =
0 \quad .
\end{equation}

Now, it is certain that a geodesic motion cannot generate GWs. If
there are also non-gravitational interactions, the conclusion of
the non-emission of GWs is still valid. A simple proof is the
following: the kinematic elements of the  non-geodesic motions
(speeds, accelerations, time derivatives of accelerations,
\emph{etc.}) are not different from the kinematic elements of
suitable purely gravitational (geodesic) motions \cite{15}. --

\vskip0.60cm \noindent\textbf{3f.} -- Consider a continuous
``cloud of dust'', described by a material energy-tensor
$T^{\mu\nu}= \varrho v^{\mu} v^{\nu}$, where $\varrho$ is the
invariant mass density and $v^{\mu}$ the four-velocity of the
``dust'' elements. A thin spacetime tube of world lines represents
the motion of a ``dust'' corpuscle. Now, as it is known, (see,
\emph{e.g.}, \cite{16}), this motion follows a \emph{geodesic}
trajectory. Consequently, \emph{no GW is emitted}. With a suitable
choice of the reference frame, it is possible to prove that also
the motions of the corpuscles of an electrically charged ``dust''
can be geodesically represented; nay, this result can be extended
to the corpuscles of a ``dust'' which interact through any field
of force \cite{17}. --

\vskip0.60cm \noindent\textbf{3g.} -- Weyl \cite{18} emphasized
that \emph{in GR} the structure of the four-dimensional world and
the concept of reference frame have acquired a characteristic
``plasticity''; as a consequence, a system of bodies in relative
motions -- \emph{e.g.}, the two stars of a binary -- can be always
reduced to rest with suitably chosen coordinate transformations.
this fact gives a trenchant proof of the non-existence of physical
GWs: indeed, their existence ought to be \emph{frame-independent}.
--

\vskip0.60cm \noindent\textbf{3h.} -- The \emph{wave-like nature}
of undulatory metric tensors (with $R_{jklm}\neq 0$, of course)
\emph{depends} on the reference system, \emph{i.e.} is only a
mathematical property of particular frames. Accordingly, these
metric tensors do not represent physical GWs \cite{19}. We have
here another precise consequence of the \emph{general covariance}
of GR. --

\vskip0.60cm \noindent\textbf{3i.} -- The Einstein-Infeld-Hoffmann
(EIH) method \cite{20} is a perturbative treatment of Einstein
equations: one develops all the functions $\varphi$ that appear in
these equations into a power series of a small parameter
$\lambda$:

\begin{equation} \label{eq:sixteen}
\varphi(x,\lambda) =  \sum_{l=0}^{\infty}  \lambda^{l} \,
\underset{l}{\varphi}(x) \quad ;
\end{equation}

thus, in particular:

\begin{equation} \label{eq:seventeen}
g_{\mu \nu} = \eta_{\mu \nu} + \sum_{n=1}^{\infty}  \lambda^{n} \,
\underset{n}{h}\quad .
\end{equation}

To determine the motions of the point-masses of a discretized
``dust'' we have two approaches at our disposal: \emph{i}) in the
original EIH approach one searches the solutions of $R_{\mu
\nu}=0$, in perfect analogy with the mass-point solutions of
Laplace equation $\Delta U=0$; \emph{ii}) in Infeld's approach one
searches the solutions of

\begin{equation} \label{eq:eighteen}
R_{\mu \nu} \, \sqrt{-g} = -8\pi \left( T_{\mu \nu} \, \sqrt{-g} -
\frac{1}{2} \, g_{\mu \nu} \, T \sqrt{-g} \right) \quad , \quad
(c=G=1) \quad ,
\end{equation}

where $T_{\mu \nu}$ is the energy tensor of a discretized ``dust''
(composed of a certain number of particles) expressed with a
proper employment of Dirac's delta-functions. In this second
approach it is assumed that the world lines of the mass-points
never intersect.

\par In the approximations higher than the second there are terms
describing a gravitational-radiation damping. However, \emph{at
any stage} one can perform an appropriate coordinate
transformation which reduces them to zero;  the equations of
motion acquire a ``Newton-like'' form \cite{21}. This result is
conceptually fundamental: it gives a significant corroboration of
the exact result of sect. \textbf{3f} concerning the purely
gravitational motions.

\vskip0.60cm \noindent\textbf{3j.} -- A famous thesis by Lorentz
and Levi-Civita (which has been formally proved \cite{22}) affirms
that in Einstein field equations the material energy-tensor
$T_{\mu \nu}$ is ``balanced exactly'' by $\left[R_{\mu \nu} -
(1/2) g_{\mu \nu} \, R\right] /\kappa $, \emph{which is the true
gravitational energy-tensor}.

\par As Levi-Civita \cite{23} emphasized, these facts have a
momentous consequence: free waves and other purely gravitational
phenomena are excluded. When the matter tensor $T_{\mu \nu}$
vanishes, the same must happen to the gravitational energy-tensor
$\left[R_{\mu \nu} - (1/2) g_{\mu \nu} \, R\right] /\kappa $.
``This fact entails a total absence of stresses, of energy flow,
and also of a simple localization of energy.'' \cite{23}. --

\vskip0.60cm \noindent\textbf{3k.} -- Levi-Civita \cite{23}
emphasized the pertinent analogy of d'Alembert Principle of
classical dynamics, which asserts that the directly applied forces
and the ``lost'' forces balance each other, with the perfect
balance  between matter tensor $T_{\mu \nu}$ and gravitational
tensor $\left[R_{\mu \nu} - (1/2) g_{\mu \nu} \, R\right] /\kappa
$, which characterizes Einstein field equations. The fact that
\emph{each} side of these equations has a \emph{zero} covariant
divergence for \emph{all} values of the metric tensor $g_{\mu
\nu}$ implies a fundamental consequence: \emph{there is no
transfer of energy, momentum, stress between matter tensor and
gravitational tensor}. Therefore, in accordance with the motions
of the bodies of a Newtonian system (cf. sect. \textbf{3i}), the
general-relativistic motions of masses do \emph{not} generate any
gravitational radiation.

\par A final remark. If we consider, \emph{e.g.}, a physical system
composed of two interacting classical fields $\Phi(x)$ and
$\Psi(x)$, moving -- for simplicity -- in a ``rigid'' spacetime
manifold (Minkowskian, or pseudo-Riemannian), the divergence of
the \emph{sum} $T_{\mu \nu}[\Phi] + T_{\mu \nu}[\Psi]$ of their
energy tensors is equal to zero. The field $g_{\mu \nu}(x)$ of GR
is a peculiar field: it represents a ``plastic'' spacetime
$\emph{\textbf{S}}$, whose ``plasticity'' is related to the fields
in the matter tensor, which move through $\emph{\textbf{S}}$
(their creation) in a ``natural'' way, without the generation of
GWs. The undulatory matric tensors, which are formal solutions of
$R_{\mu \nu}=0$, do not possess a physical reality because their
true energy-tensor is equal to zero. --

\vskip1.20cm \noindent\textbf{4.} -- We think that the future
historians of physics will treat in the same manner the beliefs in
the existence of a cosmic ether, of the gravitational waves, of
the black holes.

\end{document}